%% file: QSIM2D.tex
\documentclass[aps,pra,reprint,superscriptaddress]{revtex4-1}

\usepackage{amsmath}
\usepackage{amssymb}
\usepackage{graphicx}
\usepackage{relsize}

\pdfoutput=1

\newcommand{\eq}{\begin{equation}}
\newcommand{\eeq}{\end{equation}}
\newcommand{\eqarray}{\begin{eqnarray}}
\newcommand{\eeqarray}{\end{eqnarray}}

\graphicspath{{./images/}}

\begin{document}

\title{2D ion crystals in radiofrequency traps for quantum simulation}

\author{Philip Richerme}
\affiliation{Physics Department, Indiana University, Bloomington IN 47405, USA}

\date{\today}

\begin{abstract}

The computational difficulty of solving fully quantum many-body spin problems is a significant obstacle to understanding the behavior of strongly correlated quantum matter.  Experimental ion-trap quantum simulation is a promising approach for studying these lattice spin models, but has so far been limited to one-dimensional systems. This work argues that such quantum simulation techniques are extendable to a 2D ion crystal confined in a radiofrequency (rf) trap. Using appropriately chosen parameters, driven ion motion due to the rf fields can be made small and will not limit the types of quantum spin models that can be experimentally encoded. The rf-driven motion is calculated to modestly reduce the stability region of a 2D crystal and must be considered when designing the 2D trap. The system will be scalable to 100+ quantum particles, far beyond the realm of classical intractability, while maintaining the traditional ion-trap strengths of individual-ion control, long quantum coherence times, and site-resolved projective spin measurements.
\end{abstract}

\maketitle

\section{Introduction}
Quantum spin models are indispensable tools for describing the complex behavior of quantum condensed matter systems. They are a universal language for characterizing quantum magnetism and the behavior of quantum systems near phase transitions \cite{SachdevBook, Sachdev2008}, and they can potentially shed light on the physics that underlie exotic new materials \cite{Balents2010} or high-$T_\text{C}$ superconductivity \cite{Anderson1987,Affleck1988}. Yet, most spin models have not been ``solved," meaning that it is not possible to write down analytic expressions describing the locations of critical points, the character of different phases, or how an arbitrary initial state will evolve in time. While numeric simulations have made substantial progress investigating specific configurations \cite{DMRGReview,Foulkes2001,Sandvik2010a}, quantum many-body problems in general become intractable beyond only a few dozen spins due to the exponential scaling of Hilbert space dimension with system size \cite{Sandvik2010b}. Typically, such problems become even more computationally difficult as the dimensionality of the spin system is increased \cite{Moessner2001,Sandvik2010a,Sandvik2010b,Stoudenmire2012}.

Quantum simulation, in which the many-body problem of interest is encoded within a well-controlled experimental quantum system \cite{Feynman1982,Blatt2012}, has proven an increasingly powerful technique for studying the behavior of interacting quantum spins. Such quantum simulators should be easily reconfigurable and contain widely tunable parameters, so that they may investigate a broad variety of problems in disparate physical regimes \cite{Georgescu2014}. Recent advances have used collections of trapped ions to investigate quantum phase transitions \cite{Johanning2009,Kim2010,Islam2011,Islam2013,Richerme2013a,Richerme2013b}, explore open quantum systems \cite{Barreiro2011,Gessner2014,Kienzler2014}, witness the growth of quantum correlations and entanglement \cite{Richerme2014,Jurcevic2014}, and directly measure the many-body energy spectrum \cite{Senko2014,Jurcevic2015} in systems of up to $\sim20$ fully-coupled spins. 

In all cases, however, experiments have been restricted to one-dimensional ion chains emulating one-dimensional spin models. Although effective 2D systems can be realized in a 1D chain by applying appropriate decoupling pulses and Trotterized sequences \cite{Lanyon2011}, this approach scales very poorly due to the large number of required quantum gate operations. By instead constructing an ion trap quantum simulator with native 2D interactions, one can begin to address many of the open topics in quantum many-body physics that become important in two dimensions, such as geometric frustration \cite{DiepBook,Lhuillier2005}, exotic phases of matter (such as spin glasses \cite{Alet2006} and liquids \cite{Balents2010}), and the relationship between entanglement, frustration, and high-$T_\text{C}$ superconductivity \cite{Moessner2001,Balents2010,Anderson1973,Anderson1987}.

It would be strongly desirable for such a 2D trapped-ion quantum simulator to retain the traditional 1D ion-trap strengths: full control at the single-particle level, site-resolved measurements and readout, and spin-spin coupling rates that are fast compared to the decoherence rate. However, current efforts to build 2D trapped-ion systems in Penning traps \cite{Britton2012} and microfabricated arrays \cite{Chiaverini2008,Wilson2014} have yet to solve issues of individual ion addressing and slow coupling rates, respectively. Here, I propose the use of standard radio-frequency (rf) Paul traps for use in 2D quantum simulation experiments. I will show that it is possible to choose appropriate trap parameters so that the ions' driven motion -- called micromotion -- will have a negligible effect on the outcome of a quantum simulation, even for hundreds of trapped ions. With such parameters, the ions will remain individually addressable and resolvable, with spin-spin coupling rates comparable to those seen in 1D experiments.

The paper is organized as follows: Sec. II reviews the standard rf Paul trap, and introduces a choice of trap parameters that leads to a 2D triangular lattice of ions. Sec. III explicitly investigates the effects of micromotion on the crystal described in Sec. II, calculating the shift in equilibrium ion positions, the change in normal mode structure, and an updated stability criterion for achieving a 2D structure within an rf trap. Having developed this full analysis including micromotion, Sec. IV shows how the crystal can be used to perform quantum simulations of 2D spin models with strong coupling rates. Section V offers an outlook for future experiments and some concluding remarks.

\section{2D Paul Traps}
An rf Paul trap may be operated in a regime such that the trapped ion Coulomb crystal self-assembles in a 2D plane. We will investigate this regime here within linear Paul traps with four segmented blades (one of the most common trap designs \cite{Gulde2003,Wubbena2012,Jurcevic2014,Hucul2015,Ballance2015}), though the results are equally applicable for other trap geometries \cite{NISTPlanarTrap,Yoshimura2014}. For a typical ``blade''-style trap, the central segments of two opposing blades are driven with an rf voltage $V_0$ at frequency $\Omega_t$, while the other two central segments are held at rf ground. The outer electrode segments are biased with a dc voltage $U_0$. Near the center of the trap, the potential can be written as \cite{Wineland1998a}:
\begin{equation}
\label{eq:trappotential}
V(x,y,z,t)=\frac{V_0 \cos(\Omega_t t)}{2d_0^2} (x^2-y^2)+\frac{\kappa U_0}{2 z_0^2} (2z^2-x^2-y^2)
\end{equation}
where $d_0$ and $z_0$ are the radial and axial trap dimensions and $\kappa$ is a geometric factor of order unity. When cooled to milliKelvin temperatures, ions trapped in this potential behave as though they were in a 3D harmonic pseudopotential,
\begin{equation}
\Phi (x,y,z)=\frac{1}{2}m(\omega_x^2+\omega_y^2+\omega_z^2)
\end{equation}
where the radial and axial trapping frequencies are given by:
\begin{eqnarray}
\label{eqn:trapfreqs}
\omega_r:=\omega_x=\omega_y&=&\sqrt{\frac{Q}{m}\left(\frac{q V_0}{4d_0^2}-\frac{\kappa U_0}{z_0^2}\right)}\\
\nonumber
\omega_z&=&\sqrt{\frac{Q}{m}\frac{2\kappa U_0}{z_0^2}}
\end{eqnarray}
with $Q$ the ion charge, $m$ the ion mass, and $q \equiv 2QV_0/md_0^2\Omega_t^2$ the Mathieu ``$q$" parameter. Typically, a small asymmetry is introduced in the electrode structure to break the degeneracy of the $x$ and $y$ axes, thereby preventing a zero-frequency rotational mode and providing a unique minimum-energy configuration; for the numeric simulations to follow, the trap frequency asymmetry between the $x$ and $y$ directions is set to be $0.2\%$.


If the radial trap frequencies $\omega_r$ are much stronger than the axial frequency $\omega_z$ (the typical regime for most experiments), the ions will form a 1D chain along the central trap axis. As $\omega_z$ is increased (while holding $\omega_r$ and the number of ions fixed), the linear chain passes through a series structural phase transitions into new configurations: zig-zag, helical, and ultimately a 2D triangular lattice in the radial plane \cite{Dubin1993}. This final arrangement, which is desired for the 2D quantum simulation experiments proposed here, requires that $\omega_z$  be large compared with $\omega_r$, with the ratio scaling weakly with the number of ions $N$ in the trap \cite{Dubin1993}:
\begin{equation}
\label{eq:anisotropy}
\frac{\omega_z}{\omega_r} > (2.264 N)^{1/4}
\end{equation}
Hence, a 100-ion system requires an axial frequency $\gtrsim 4$ times larger than the radial frequencies. From Eq. \ref{eqn:trapfreqs}, a large axial frequency can be achieved by increasing the voltages $U_0$ applied to the outer trap segments. However, if the dc voltage $U_0$ is too large compared with the rf voltage $V_0$, this has a destabilizing effect on the crystal: the radial frequency (Eq. \ref{eqn:trapfreqs}) becomes imaginary, and ions no longer have bound trajectories. The ion trap voltages, frequencies, and sizes thus must be carefully chosen in order to achieve a stable and robust 2D planar crystal. The full stability regime for a 2D ion lattice in an rf trap, which depends on the axial and radial trap frequencies as well as the number of ions, will be calculated and shown in Sec. III(c).

For the numeric simulations presented in upcoming sections, I choose the following set of trap parameters: $d_0=z_0=200~\mu\text{m},~\Omega_t=(2\pi)\times 50~ \text{MHz},~V_0=440~ \text{V},$ and $\kappa U_0=13~ \text{V},$ which result in radial and axial trap frequencies of $\omega_r=(2\pi)\times 510$ kHz and $\omega_z=(2\pi)\times 3.04~ \text{MHz}$. These parameters lie squarely within the stability regime for a 2D crystal of $^{171}$Yb$^+$ ions and are all straightforward to achieve in the laboratory. However, before implementing such a trap for quantum simulation experiments, we must first consider the effects of rf driven motion on the 2D ion crystal.

\section{Effects of Micromotion}
For a 2D crystal in a linear Paul trap, each ion is subject to rf-driven micromotion with an amplitude proportional to the ion's distance away from the central trap axis. For several types of ion-trap experiments, even small amounts of micromotion can have significant harmful effects: it can lead to large systematic Doppler shifts in ion-based atomic clocks \cite{Berkeland1998,Keller2015}, and it can substantially reduce the fidelities of quantum gates during a quantum computation (in the absence of advanced micromotion-correcting protocols) \cite{Shen2014,Wang2015}. Nevertheless, many successful experiments do not depend sensitively on micromotion amplitude \cite{Mortensen2006,Ostendorf2006,Tong2010,Schwarz2012,Tabor2012}, and ion Coulomb crystals of up to $10^5-10^6$ particles have been confined in rf Paul traps \cite{Itano1998,Drewsen1998}.

\begin{figure*}[t!]
\begin{center}
\includegraphics*[width=.5\textwidth]{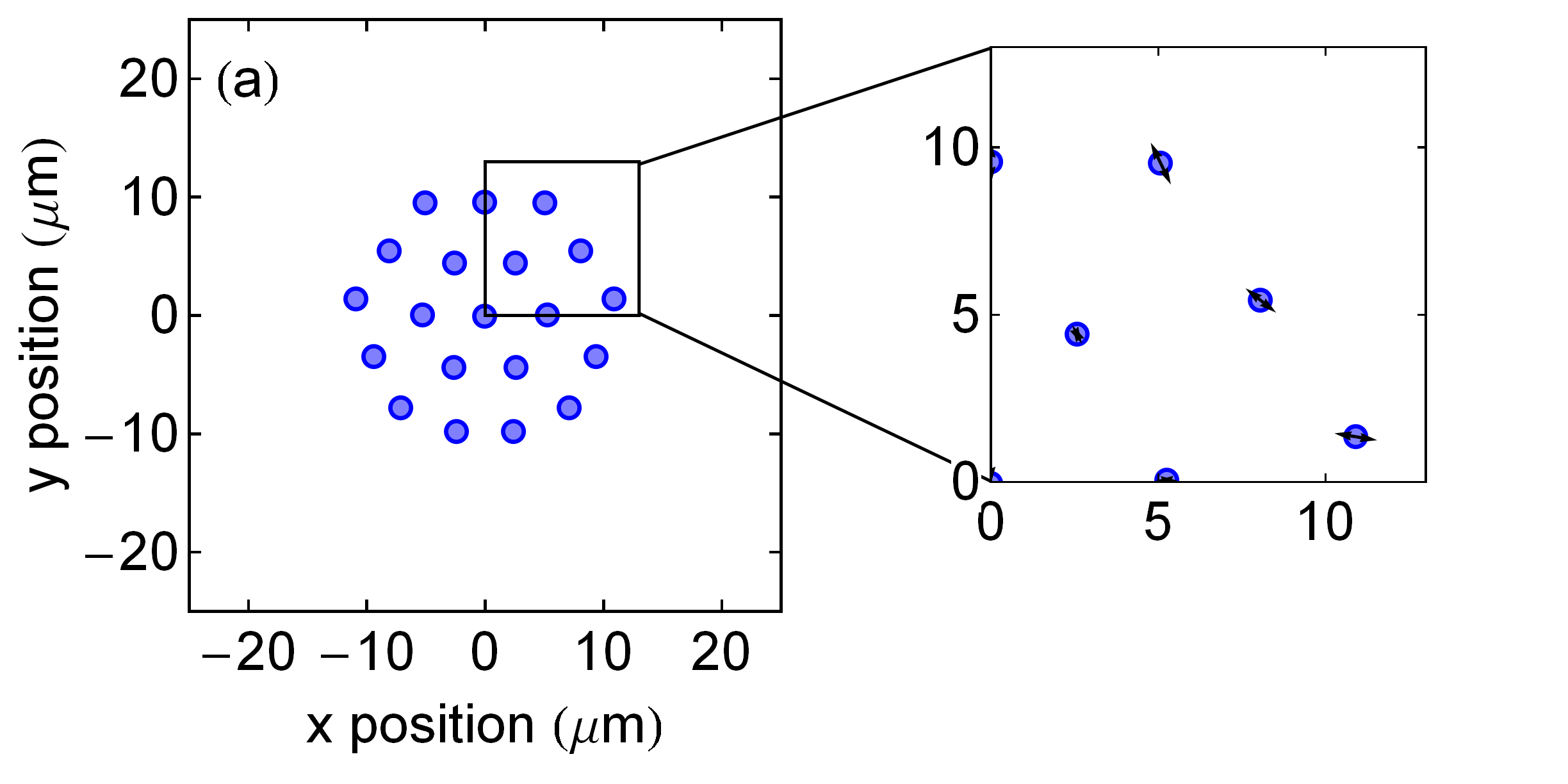}\includegraphics*[width=.5\textwidth]{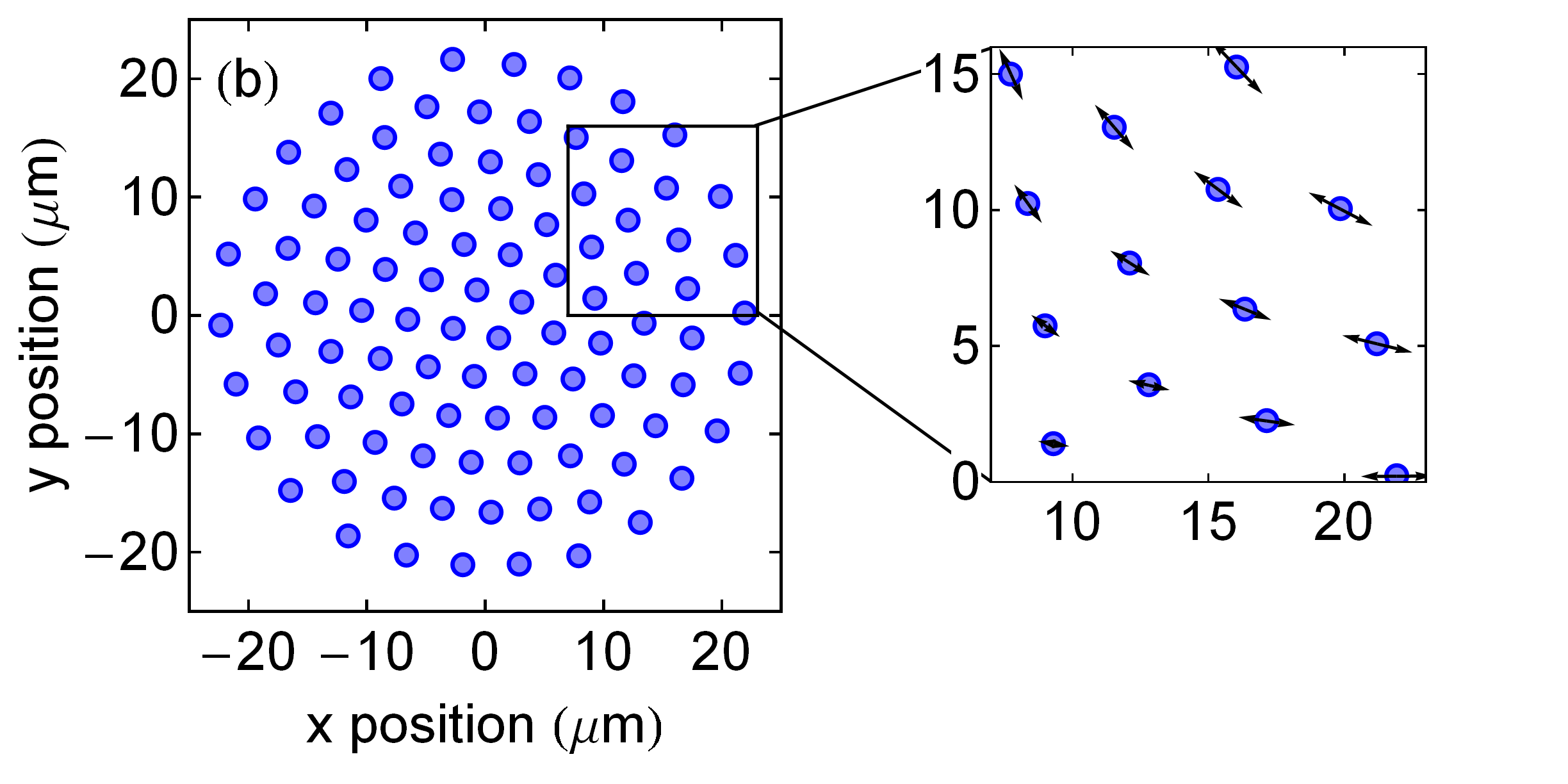}
\end{center}
\caption{Equilibrium ion positions in the proposed rf Paul trap for (a) 20 ions and (b) 100 ions. Results were obtained with a molecular dynamics simulation that included the effects of micromotion. The insets show the micromotion directions and amplitudes, which are small compared to the inter-ion spacing.}
\label{fig:micromotion}
\end{figure*}

This section will demonstrate that for carefully chosen parameters (such as those introduced above), micromotion effects on 2D quantum simulation experiments are both predictable and small. This result is enabled by the orientation of the 2D crystal: there is effectively no micromotion amplitude in the axial (transverse) direction, since the ions are compressed to a single plane at the axial trap center. By utilizing the axial modes of motion for quantum simulation protocols, one can thus sidestep the large effects of micromotion in the radial plane. Nevertheless, 4 residual effects in the axial plane must still be considered: (1) a shift in the equilibrium ion positions, (2) the resolvability of ions due to radial micromotion amplitude, (3) shifts in the axial normal mode frequencies, and (4) an altered stability region for maintaining a 2D crystal. These effects are each explored in detail below.

\subsection{Equilibrium Positions and Micromotion Amplitude}

In the absence of micromotion, finding the equilibrium positions for a 2D ion crystal proceeds similarly to the 1D case. In the radial plane, the potential experienced by the ions has contributions from the trap voltages as well as the Coulomb interaction:
\begin{equation}
\begin{split}
V(x,y) & =\sum_i \left( \frac{1}{2}m\omega_x^2 x_i^2 + \frac{1}{2}m\omega_y^2 y_i^2 \right) \\
& +\sum_{i<j}\frac{e^2}{4\pi\epsilon_0 \sqrt{(x_i-x_j)^2+(y_i-y_j)^2}}
\end{split}
\end{equation}
The equilibrium positions are the set of coordinates $\{x_i,y_i\}$ that minimize the energy. Although direct numerical minimization is possible for small system sizes, finding the equilibrium configuration for larger numbers often requires the use of molecular dynamics simulations with added dissipation \cite{Schiffer2000}; this is the approached used for calculations here.

In the presence of micromotion, there is no longer an ``equilibrium'' position; the radial coordinates of each ion vary in time as
\begin{equation}
\label{eq:mmposition}
\vec{r}(t)=\vec{r}_0+\vec{r}_1 \cos(\Omega_t t)+\vec{r}_2 \cos(2 \Omega_t t)+\ldots
\end{equation}
where $\vec{r}_0$ is the ion's average position, and the higher order terms indicate the amplitude of the motion at the $n^{\text{th}}$ harmonic of the drive frequency $\Omega_t$. Following \cite{Shen2014,Wang2015}, each of these amplitudes $\vec{r}_n$ for each ion can be extracted by self-consistently solving the equations of motion within the full trap potential of Eq. \ref{eq:trappotential}.

The calculated central positions $\vec{r}_0$ for a 20- and 100-ion crystal, using the parameters introduced in Sec. II and including micromotion, are shown in Fig. \ref{fig:micromotion}a-b. The crystals self-organize into a 2D triangular lattice, with a 4.3 $\mu$m average inter-ion spacing for the 100-particle case. When accounting for micromotion, the positions are found to shift by an average of only 0.08 $\mu$m -- a very small change that would not be noticeable with standard imaging techniques and will not have any foreseeable consequences for 2D quantum simulations.

The amplitude of the radial micromotion is also important to consider; large amplitude excursions could obscure the individual ion positions, preventing quantum spin readout. The amplitudes of the micromotion-induced terms in Eq. \ref{eq:mmposition} may be written as $|\vec{r}_1|=q/2$ and $|\vec{r}_2|=q^2/32$ \cite{Landau1976p9395,Wineland1998a}, where $q\equiv 2QV_0/md_0^2 \Omega_t^2$ is the Mathieu parameter. Since $q < 1$ for any stable rf trap, the first order amplitude dominates the time-dependent part of Eq. \ref{eq:mmposition}, and smaller $q$ parameters are advantageous for minimizing the micromotion amplitude. For the trap settings proposed in Sec. II, $q = 0.125$; for the 100-ion case, this leads to a micromotion amplitude of 1.4 $\mu$m for the most radially extended particle -- still small compared to the 4.3 $\mu$m inter-ion distance (see Fig. \ref{fig:micromotion} insets). Since the radial extent of the 2D crystal scales as $\sim d\sqrt{N}/2$ for $N$ particles with separation distance $d$, the maximum micromotion amplitude scales as $\sim qd\sqrt{N}/4$. If we demand that this amplitude be smaller than half the inter-ion distance ($d/2$), this constrains the maximum number of trapped ions to be $N \approx 4/q^2$, which is $N \approx 250$ for the chosen trap parameters.

\subsection{Normal Mode Structure}
To calculate the normal mode frequencies of an ion crystal, one typically expands the Coulomb potential around the equilibrium positions to second order, then diagonalizes the resulting matrix to find the eigenfrequencies and eigenmodes \cite{James1998}. However, the presence of micromotion in the radial plane can have a notable effect on the axial mode frequencies and eigenfunctions (even though there is no axial micromotion), and must be taken into account \cite{Wang2015}. As we will see, the primary effect of radial micromotion is to decrease the frequencies of all axial modes (except for the center-of-mass), with the low-frequency modes experiencing the largest shift.

\begin{figure}[t!]
\begin{center}
\includegraphics[width=\columnwidth]{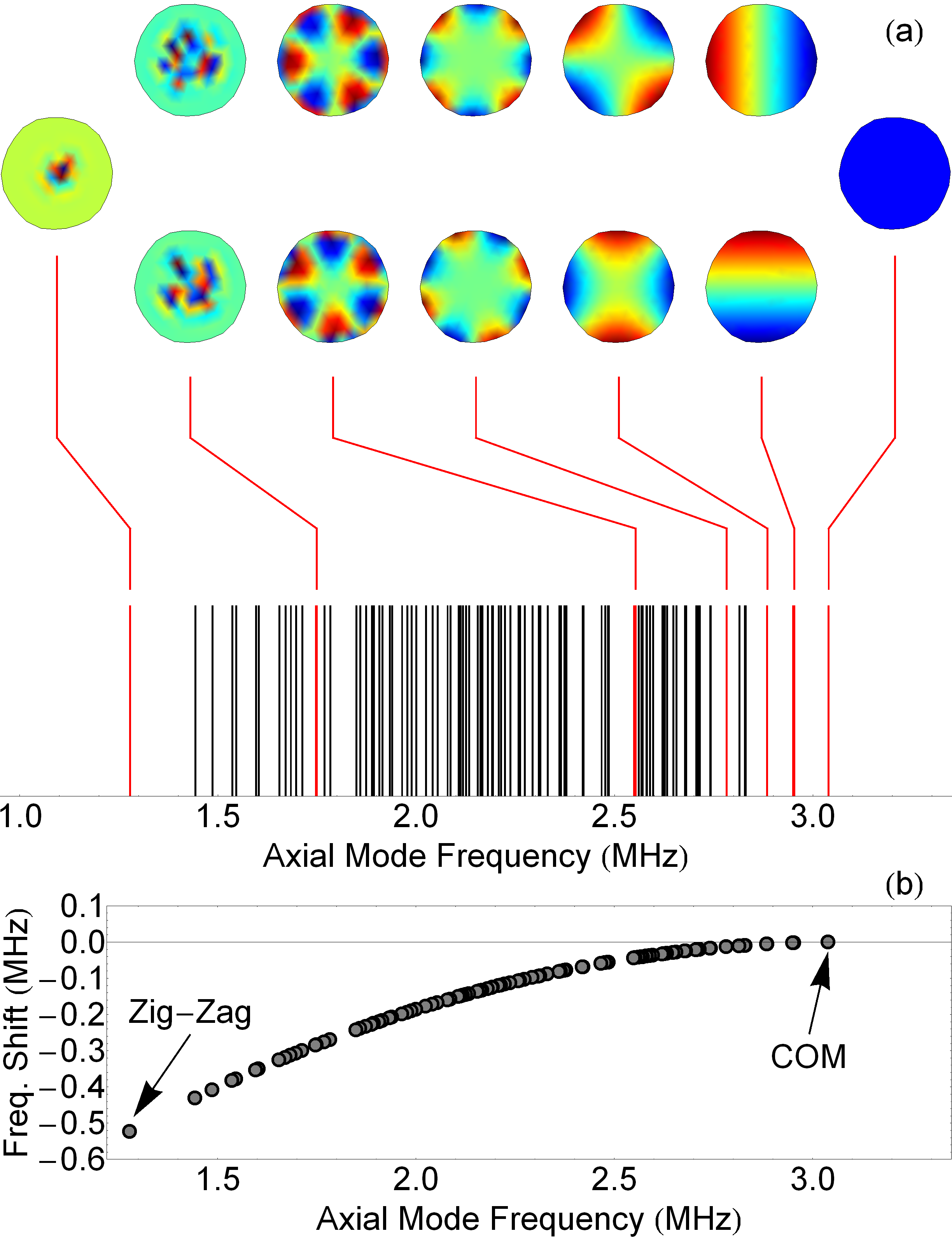}
\end{center}
\caption{(a) Calculated spectrum of axial mode frequencies (including micromotion effects), along with several mode eigenfunctions, for a 100-ion crystal using the parameters in Sec. II. The highest frequency mode is the center-of-mass motion (no spatial variation), while lower frequency modes vary on shorter and shorter length scales. (b) Radial micromotion induces a frequency shift in the axial normal modes, compared to the no-micromotion case. The center-of-mass mode is left unchanged, while the lower frequency modes are shifted downwards by progressively larger amounts.}
\label{fig:modes}
\end{figure}

To second order, the axial potential experienced by the ions is given by
\begin{equation}
\begin{split}
V(z)= & \sum_i \frac{1}{2}m\omega_z^2 z_i^2 + \frac{e^2}{4\pi\epsilon_0}\sum_{i\neq j} \left( \frac{1}{r_{ij}^3} \right) z_i z_j \\
&-\frac{e^2}{4\pi\epsilon_0}\sum_{i\neq j} \left( \frac{1}{r_{ij}^3} \right) z_i^2
\end{split}
\end{equation}
where $r_{ij}\equiv\sqrt{(x_i-x_j)^2+(y_i-y_j)^2}$. Due to the radial micromotion, the ion position differences $r_{ij}(t)$ are dynamic (Eq. \ref{eq:mmposition}), which implies that the axial normal mode frequencies will also inherit a time dependence. However, since $\Omega_t \gg \omega_z$, it is sufficient to consider the expectation value of the potential $V(z)$ over one period of micromotion, and diagonalize this resulting matrix to find the normal mode eigenfrequencies.

The results of this calculation are shown in Fig. \ref{fig:modes}(a), along with several representative mode eigenfunctions, for 100 trapped ions. Here, the highest frequency axial (transverse) motion corresponds to the center-of-mass (COM) mode (just as in the 1D case) and remains unchanged in the presence of micromotion. However, Fig. \ref{fig:modes}(b) shows that all other mode frequencies are suppressed under micromotion; for the lowest (zig-zag) mode, the reduction is more than 0.5 MHz $(\sim 30\%)$. This change in mode structure will not impede 2D quantum simulations; rather, its effect will be to increase the effective spin-spin interaction range between ions in the lattice (as will be shown in Sec. IV). When quantum simulating spin models that are sensitive to the specific value of interaction range, one must thus take into account this micromotion-induced frequency shift.

\subsection{Stability Region}
Consider a 1D linear ion chain, which is characterized by a set of transverse normal mode frequencies. If the axial frequency $\omega_z$ is increased while holding all other parameters fixed, the ions will be pushed closer together, and all transverse mode frequencies (except for the COM) will shift downwards. Eventually, the chain will undergo a structural phase transition and ``buckle'' at the center; this corresponds to the lowest transverse mode (i.e. the zig-zag mode) shifting downwards to zero frequency. 

\begin{figure}[t!]
\begin{center}
\includegraphics[width=\columnwidth]{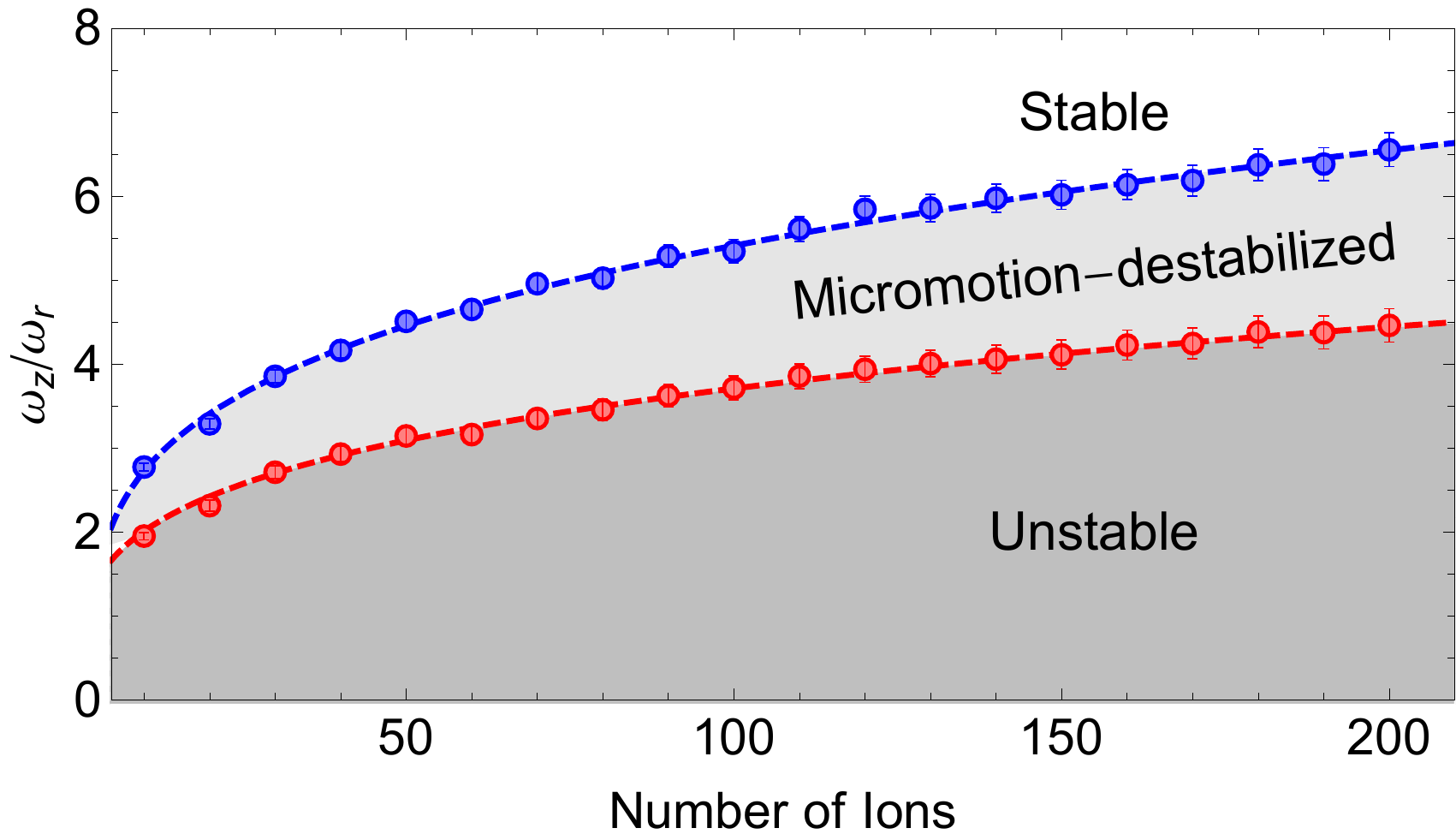}
\end{center}
\caption{Stability region for a 2D planar ion crystal, using the trap parameters of Sec. II and $\omega_r=0.5$ MHz. Red points show the calculated boundary in the absence of micromotion, in agreement with Eq. \ref{eq:anisotropy} (red dashed line). When micromotion effects are included, the entire region below the blue points becomes unstable. This stability boundary is also found to exhibit weak power-law scaling with the number of ions (blue dashed line).}
\label{fig:stability}
\end{figure}

Analogous effects occur for 2D planar ion crystals: when the radial confinement becomes too strong compared to the axial confinement, the zig-zag transverse mode crosses zero frequency, and the crystal buckles into 3D. This effect has been well-described in \cite{Dubin1993}, which led to the scaling law in Eq. \ref{eq:anisotropy}. However, this earlier analysis did not explicitly consider rf traps, where micromotion can have significant impacts on crystal stability.

The previous section (as well as Fig. \ref{fig:modes}) demonstrated that micromotion itself causes a downward shift in the transverse mode frequencies. This observation implies that micromotion can have a destabilizing effect: the zig-zag mode can be pushed to zero-frequency, even for crystals that are ``stable'' according to Eq. \ref{eq:anisotropy}. Thus, one must carefully calculate the mode structure -- including micromotion -- before concluding that the ions will remain in the desired planar geometry.

Fig. \ref{fig:stability} shows the boundary between stable and unstable 2D planar crystals, using the trap parameters of Sec. II with $\omega_r=0.5$ MHz, when micromotion is ignored (red points) and included (blue points). As argued above, the stability region including micromotion is reduced when compared with the predictions of Eq. \ref{eq:anisotropy}. As before, the boundary scales weakly with the number of ions $N$; now, $\omega_z/\omega_r \propto N^{0.27 \pm 0.01}$ (compared to $N^{0.25}$ previously). The primary effect of micromotion is thus a multiplicative increase in the ratio $\omega_z/\omega_r$ required for stability; a \emph{very} rough rule would be to calculate the needed frequency ratio via Eq. $\ref{eq:anisotropy}$, then add 45$\%$ to account for effects of micromotion.

\section{Generating 2D spin-spin interactions}
By loading $^{171}$Yb$^+$ ions into the trap described above, one can engineer an effective 2D many-body spin system for quantum simulations. As in 1D ion trap experiments, effective spin qubits can be encoded in the hyperfine ground states \cite{Enders1993,Hannemann2002,Olmschenk2007}, which are first-order insensitive to external magnetic field noise and can yield coherence times of over 10 minutes \cite{Fisk1997}. If the ions are irradiated with an appropriate frequency of laser light, the spin-dependent fluorescence from each ion allows for a high-fidelity measurement of the projected spin state \cite{Olmschenk2007,Noek2013}.

\begin{figure}[b!]
\begin{center}
\includegraphics[width=\columnwidth]{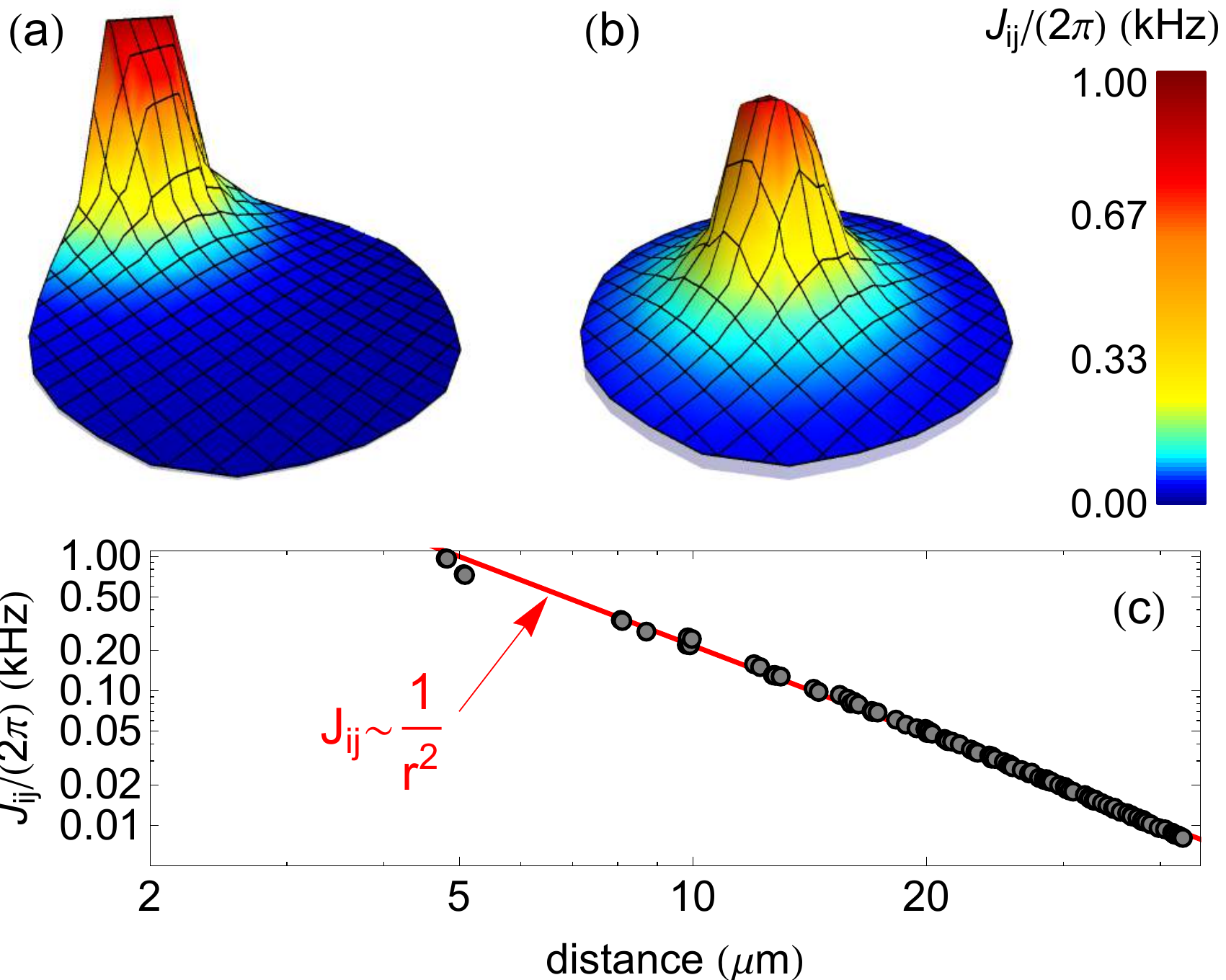}
\end{center}
\caption{The normal modes determine the spin-spin couplings $J_{ij}$ (Eq. \ref{eq:Jij}), which are shown for an edge spin (a) and a central spin (b) in a 100-ion array using the trap parameters detailed in the text. For these parameters, the couplings fall of algebraically with distance as $\sim 1/r^\alpha$, with $\alpha$ tunable between 0 and 3. Panel (c) shows this power-law decay for the edge spin chosen in (a).}
\label{fig:Jij}
\end{figure}

The effective spin qubits within each ion can be coupled together by applying spin-dependent optical dipole forces \cite{Molmer1999,Sorensen2000}. These forces are induced by global, far-detuned Raman transitions at 355 nm that virtually excite the collective modes of ion motion. If the wavevector difference between the two Raman beams $\Delta k$ lies along the axial direction of the trap, this will excite the axial motional modes $\omega_m^z$ (the desired direction to avoid problematic micromotion effects). The two beams should contain a pair of beat-note frequencies symmetrically detuned from the hyperfine splitting by a frequency $\mu$, which is comparable to the center-of-mass axial frequency. When a resonant carrier interaction is added, this arrangement results in an effective transverse-field Ising Hamiltonian \cite{Porras2004,Kim2009};
\begin{equation}
H_{Ising}=\sum_{i<j}J_{ij}\sigma_i^x\sigma_j^x+B \sum_i \sigma_i^y
\label{eq:ising}
\end{equation}
where $\hbar$ has been set to 1, $B$ is an effective transverse magnetic field, and the long-range spin-spin couplings are given by
\begin{equation}
J_{ij}=\Omega^2\frac{\hbar \Delta k^2}{2m}\sum_{m=1}^N \frac{b_{im}b_{jm}}{\mu^2-(\omega_m^z)^2}
\label{eq:Jij}
\end{equation}
where $\Omega$ is the carrier Rabi frequency, and $b_{im}$ the normal mode eigenvector component of the $i^{th}$ ion in mode $m$.

\begin{figure*}[t!]
\begin{center}
\includegraphics[width=0.24\textwidth]{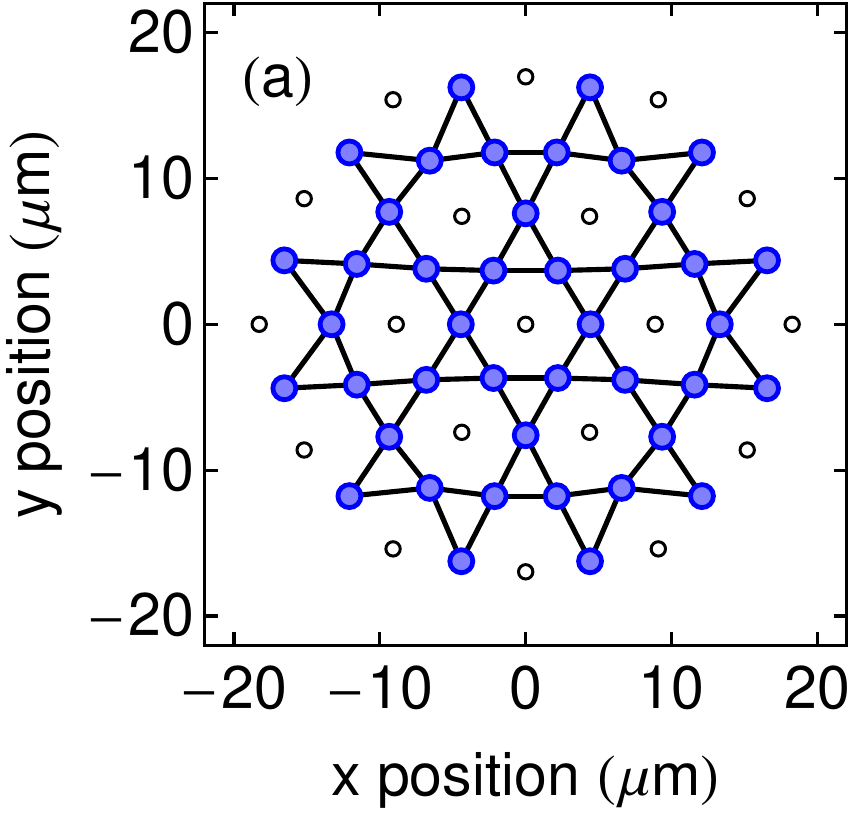}
\includegraphics[width=0.24\textwidth]{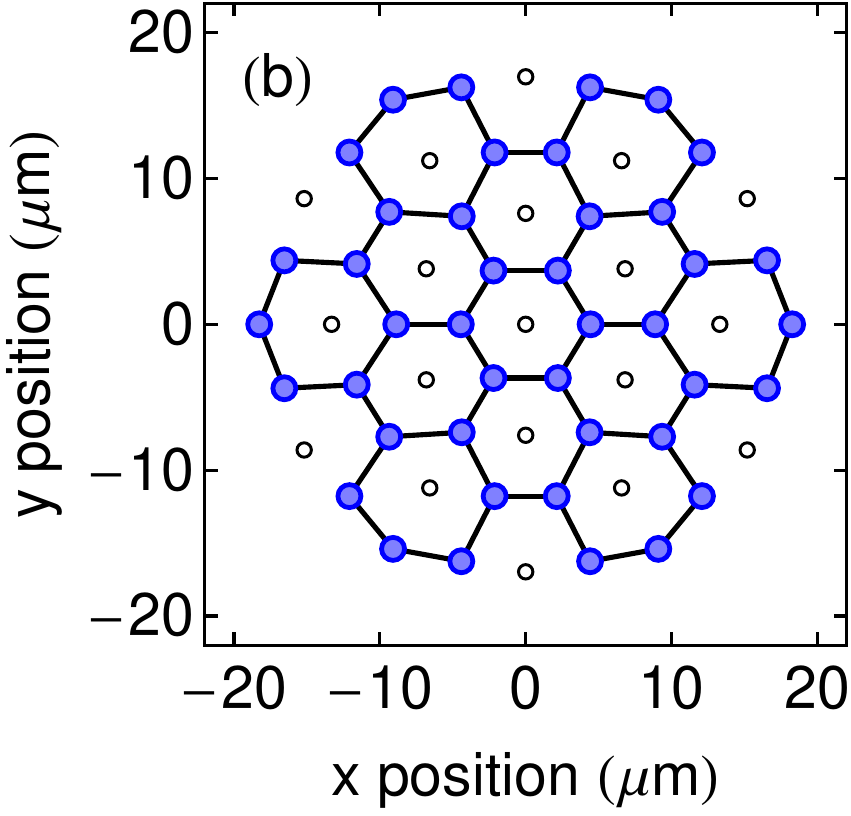}
\includegraphics[width=0.24\textwidth]{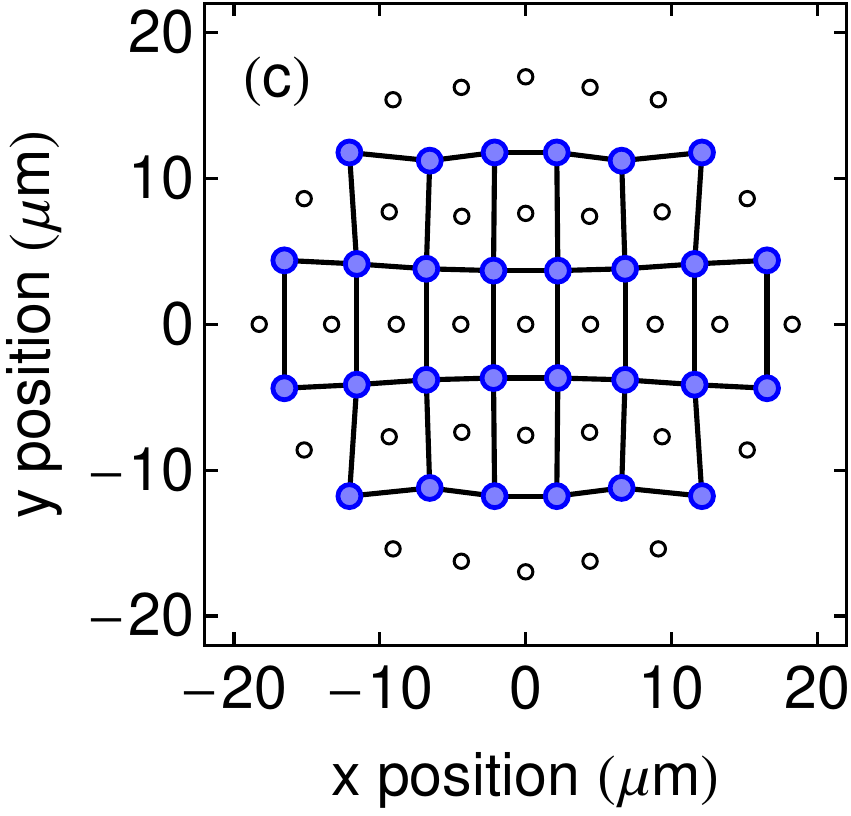}
\includegraphics[width=0.24\textwidth]{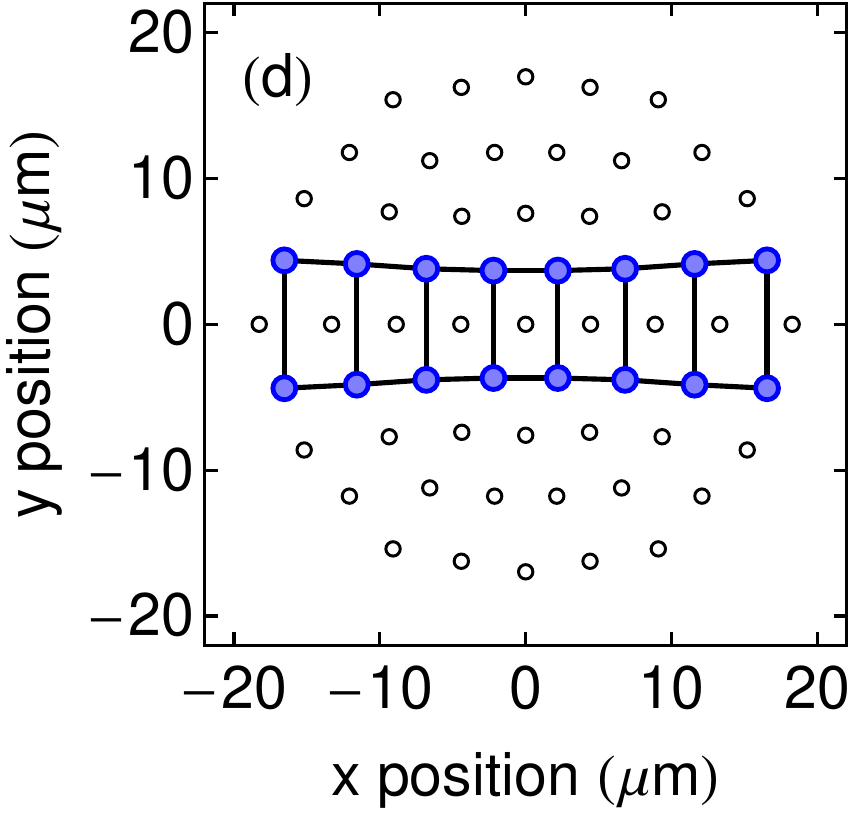}
\end{center}
\caption{By using a focused laser beam to shelve specific ions in uncoupled spin states, one can realize a variety of different lattice geometries. Four examples are depicted here: (a) Kagom\'e, (b) Honeycomb, (c) Rectangular, and (d) Spin Ladder. Blue circles are participating ions, empty circles are ``hidden" ions, and black lines indicate the nearest-neighbor spin-spin couplings on each participating lattice site.}
\label{fig:lattices}
\end{figure*}

Having calculated the mode frequencies $\omega_m^z$ and eigenfunctions $b_{im}$ using the techniques outlined in Sec. III, the complete spin-spin coupling matrix $J_{ij}$ follows immediately (Eq. \ref{eq:Jij}), and is shown in Fig. \ref{fig:Jij}(a)-(b) for two different spins in the 2D lattice. Here, the carrier Rabi frequency is set to $\Omega=(2\pi)\times 1.5$ MHz, and the laser detuning $\mu$ was chosen to be blue of the center-of-mass axial mode by approximately $3\Omega\sqrt{\hbar\Delta k^2/2m\omega_z}=(2\pi)\times 350$ kHz (ensuring that residual phonon effects are kept small \cite{Islam2011}). These proposed parameters yield a nearest-neighbor antiferromagnetic spin-spin coupling of approximately $(2\pi)\times 1$ kHz (which is nearly an order of magnitude faster than typical decoherence rates \cite{Richerme2014,Jurcevic2014}), as well as a long-range coupling that decays with distance $r$ as $\sim 1/r^2$. In general, by choosing different values of $\mu$, long-range interactions can be continuously tuned to decay with distance as any power between $1/r^0$ and $1/r^3$ \cite{Kim2009}.

\section{Conclusions}

This work has argued that rf Paul traps with appropriately chosen parameters can serve as a scalable platform for developing 2D quantum simulation experiments. In such traps, the ions self-assemble into a triangular lattice with tunable, long-range couplings given by Eq. \ref{eq:Jij}. As pictured in Fig. \ref{fig:lattices}, it would also be possible to shelve specific ions in electronic states outside of the qubit subspace, allowing for multiple types of effective lattice configurations \cite{Bermudez2012}. The residual effects of micromotion in 2D traps can be well-characterized, leading to a slightly longer spin-spin interaction range and a reduced (but still easily achievable) 2D trap stability region. It should be straightforward to achieve several hundreds of ions for use in these experiments, limited by the ability to individually resolve and address the atoms (see Sec. III(a)) and the ability to stably confine large numbers in a 2D configuration using reasonable laboratory voltages (see Fig. \ref{fig:stability}).

Using an appropriate combination of Raman laser frequencies, amplitudes, and phases, it will be possible to quantum simulate a diverse toolbox of spin-model Hamiltonians. Ising couplings naturally occur as a result of the applied spin-dependent force \cite{Porras2004}, and have been demonstrated in numerous 1D quantum simulation experiments \cite{Kim2010,Islam2011,Richerme2013a,Richerme2013b,Islam2013,Senko2014}. Recently, 1D simulations have also demonstrated an XY spin model by applying a large resonant Raman transition in conjunction with the Ising interaction \cite{Richerme2014,Jurcevic2014}. While not yet experimentally demonstrated, it should also be possible to realize full Heisenberg spin-spin interactions of the form $H=\sum J_{ij} \vec{\sigma_i}\cdot\vec{\sigma_j}$, following the ideas proposed in Refs. \cite{Cohen2015,Bermudez2016Unpublished}. These 1D advances can be directly applied to 2D quantum simulations, since they both fundamentally operate by coupling $^{171}$Yb$^+$ ions together through collective motional modes. With the ability to apply a variety of different spin Hamiltonians, along with tunable spin-spin coupling strengths $J_{ij}$, one can fully implement a quantum simulator to explore many of the important open questions in 2D quantum many-body physics.

\section{Acknowledgments}
The author is grateful to Shengtao Wang for illuminating discussions. This work is supported by the Air Force Office of Scientific Research award no. FA9550-16-1-0277, and the College of Arts and Sciences at Indiana University.

\input{QSIM2D.bbl}

\end{document}

%% file: QSIM2D.bbl
%